\documentstyle[prl,aps]{revtex}
\begin{document}

\title{QUARK-GLUON PLASMA\footnote{Presented at the XXXVIII Cracow 
School of Theoretical Physics, Zakopane, Poland, June 1-10, 1998; 
Acta Phys. Pol. {\bf B29} (1998) 3711.}}

\author{Stanis\l aw MR\'OWCZY\'NSKI\footnote{E-mail: 
{\tt mrow@fuw.edu.pl.}}}

\address{So\l tan Institute for Nuclear Studies \\
ul. Ho\.za 69, PL - 00-681 Warsaw, Poland \\
and Institute of Physics, Pedagogical University \\
ul. Konopnickiej 15, PL - 25-406 Kielce, Poland}

\maketitle

\begin{abstract}

An elementary introduction to the physics of quark-gluon plasma is given. 
We start with a sketchy presentation of the Quantum Chromodynamics which 
is the fundamental theory of strong interactions. The structure of hadrons 
built up of quarks and gluons is briefly discussed with a special emphasis 
on the confinement hypothesis. Then, we explain what is the quark-gluon 
plasma and consider why and when the hadrons can dissolve liberating the 
quarks and gluons. The heavy-ion collisions at high-energies, which provide 
a unique opportunity to get a droplet of the quark-gluon plasma in the 
terrestrial conditions, are described. We also consider the most promising 
experimental signatures of the quark-gluon plasma produced in nucleus-nucleus 
collisions. At the end, the perspectives of the quark-gluon plasma studies 
at the future accelerators are mentioned.

\end{abstract}
\vspace{1cm}

\section{Introduction}

The quark-gluon plasma is a state of the extremely dense matter with 
the quarks and gluons being its constituents. Soon after the Big Bang 
the matter was just in such a phase. When the Universe was expanding 
and cooling down the quark-gluon plasma turned into hadrons - neutrons 
and protons, in particular -  which further formed the atomic nuclei. 
The plasma is not directly observed in Nature nowadays, but the astrophysical 
compact objects as the neutron stars may conceal the quark-gluon nuggets 
in their dense centers. The most exciting however are the prospects to study 
the plasma in the laboratory experiments. A broad research program of 
the heavy-ion collisions, which provide a unique opportunity to produce 
the quark-gluon droplets in the terrestrial conditions, is underway. 
While the question, whether the plasma is already produced with 
the currently operating heavy-ion accelerators, is right now vigorously 
debated, there are hardly any doubts that we will have a reliable evidence 
of the plasma within a few years. Then, there will be completed the 
accelerators of a new generation: the Relativistic Heavy Ion Collider 
(RHIC) at the Brookhaven National Laboratory and the Large Hadron Collider 
(LHC) at CERN.

The aim of this article is to give a very elementary introduction to the 
physics of quark-gluon plasma. We start with a few words on the Quantum 
Chromodynamics which is a fundamental theory of strong interactions. 
Then, the structure of hadrons, which are built up of quarks and gluons, 
is briefly discussed. A particular emphasis is paid on the confinement 
hypothesis. In the next section we try to explain why and when the 
hadrons can dissolve liberating the quarks and gluons from their interiors. 
Finally, the generation of the quark-gluon plasma in heavy-ion collisions 
is considered.

There is a huge literature on every topic touched in this article. 
Instead of citing numerous original papers, we rather recommend two 
collections of the review articles \cite{Hwa90,Hwa95}. The more recent 
progress in the field can be followed due to the proceedings of the
regular ``Quark Matter" conferences \cite{Pro93,Pro95,Pro96}.

Throughout the article we use the natural units where the velocity of 
light $c$, the Planck $\hbar$ and Boltzmann $k$ constants 
are all equal unity. Then, the mass, momentum and temperature have the 
dimension of energy, which is most often expressed in MeV. The distances 
in space or time are either measured in the length units, which are 
usually given in fm $( 1\;{\rm fm} = 10^{-13}\:{\rm cm})$, or in the 
inverse energy units. One easily recalculates the length into the 
inverse energy or vice verse keeping in mind that 
$\hbar c = 197.3 \;\; {\rm MeV\!\cdot fm}$.

\section{Quantum Chromodynamics}

Quantum Chromodynamics (QCD) strongly resembles the Quantum 
Electrodynamics (QED). While QED describes the interaction of electric 
charges - usually electrons and their antiparticles positrons - with the 
electromagnetic field represented by photons, QCD deals with the 
quarks and gluons corresponding to, respectively, the electrons and 
photons. The quarks are, as electrons, massive and curry a specific 
charge called color, which is however not of one but of three types: 
red, blue and green. The gluons, which are, as photons, massless, 
but not neutral. In contrast to photons, they curry color charges being 
the combinations of the quark ones. The electromagnetic interaction 
proceeds due to the photon exchanges. Analogously the quarks 
interact exchanging the gluons. Although the photons being neutral 
cannot interact directly with each other, there are forces acting 
between gluons.
  
QCD has emerged as a theory of quarks and gluons which built up 
the hadrons i.e. strongly interacting particles such as neutrons and 
protons. However, there is no commonly accepted model of the hadron 
structure. The difficulty lies in the very nature of the strong interaction - 
its strength. While the perturbative expansion, where the noninteracting 
system is treated as a first approximation, appear to be the only effective 
and universal computational method in the quantum field theory, a large 
value of the QCD coupling constant  excludes applicability of the method 
for the system of quarks and gluons. However, QCD possesses a remarkable 
property called the {\it asymptotic freedom}. The coupling constant $\alpha_s$ 
effectively depends on the four-momentum $Q$ transferred in the interaction as
\begin{equation}\label{alpha}
\alpha_s(Q) = {12 \pi \over (33 - 2N_f ) \, {\rm ln}Q^2/\Lambda_{QCD}^2} \;,
\end{equation}
where $N_f$ is the number of the number of the quark flavours (types) and 
$\Lambda_{QCD}$ is the QCD scale parameter, $\Lambda_{QCD} \cong 200$ 
MeV. Eq. (\ref{alpha}) shows that the coupling constant is small when 
$Q^2 \gg \Lambda_{QCD}^2$. Therefore, the interactions with a large 
momentum transfer can be treated in the perturbative way. QCD appears to 
be indeed very successful in describing the hard processes, such as a 
production of jets in high energy proton-antiproton collisions, which 
proceed with high $Q$. 

\section{Hadron structure}

The description of the soft processes in QCD, which, in particular, control 
the hadron structure, remains a very serious unresolved  problem of the strong
interactions. One has to rely on the phenomenological models, the validity of
which can be only tested by confrontation against the experimental data. 
Dealing with the soft QCD one often uses the concept of the constituent 
quarks which should be distinguished the current quarks. The latter ones are 
the fundamental elementary spin 1/2 particles which are present in the QCD
lagrangian. The current up ($u$) and down ($d$) quarks are light with the 
mass of a few MeV. The constituent $u$ and $d$ quarks are supposed to be 
the effective quasiparticles with the (large) masses of  about 300 MeV 
(1/3 of the nucleon) generated by the interaction. The difference between 
the current and constituent strange ($s$) quark masses is less dramatic. 
They are, respectively, about 150 and 450 MeV. In the case of heavy 
quarks - charm ($c$), bottom ($b$) and top ($t$) - the distinction is no 
longer valid.

Within the constituent quark model the baryon is described as three 
bounded quarks. The meson is then the system of quark and antiquark. 
In terms of current quarks the hadron is seen as a cloud of quarks 
and gluons. The baryon is then no longer built of three quarks and the 
meson of quark and antiquark. Instead, the hadron curries the quantum 
numbers of, respectively, three quarks or the quark and antiquark. 
Therefore, the hadron is composed of the valence quarks (quark and 
antiquark in the case of meson and three quarks for the baryon) and 
the sea constituted by the quark-antiquark pairs and gluons. 

The gluons are believed to glue together the quarks which form
the hadron, but a satisfactory theory of the hadron binding is still 
missing. Such a theory has to explain the {\it hypothesis of confinement} 
which is the fundamental element of our understanding of the hadron 
world. While the electric charges tend to form electrically neutral atoms 
and molecules, in the case of the chromodynamic interactions there 
seems to be a strict rule that the color charges occur only within the 
white configurations. The existence of  the separated color objects as 
quarks and gluons is excluded. They must be confined in the colorless 
systems such as hadrons. The three quarks forming a baryon curry three 
fundamental colors which give all together the white object. In the 
meson case the quark color is complementary to the color of the antiquark. 
In principle, the confinement hypothesis allows for the existence of not 
only the meson (quark-antiquark) and baryon (three quark) configurations 
but for any white one such as a dibaryon which is the six quark system. 
However, in spite of hard experimental efforts, the reliable evidence for 
the hadrons different than baryons and mesons is lacking.

There are many phenomenological approaches to the confinement. Let us 
briefly present here the string model inspired by the Meissner effect which 
is the expelling of the magnetic field from the superconducting material. 
The model assumes that the vacuum behaves as a diaelectric medium,
where the chromodynamic field cannot propagate but is confined in thin
tubes or strings which connect the field sources. In Fig. 1 we show the 
electric field generated by the two opposite charges which are in the (normal) 
vacuum (a) and in the diaelectric medium (b). Let us compute the potential 
which acts between charges in the latter case. Using the Gauss theorem one 
immediately finds the electric field as $E = q /\sigma $, where $q$ denotes 
the charge and $\sigma $ the cross section of the tube. If  $\sigma$ is 
independent of  the distance $r$ between the charges, their potential energy 
equals 
\begin{equation}\label{linear}
V(r) = {q^2 \over \sigma} \, r \;.
\end{equation}
As seen the potential energy grows lineary with $r$ when the charges are 
put in the diaelectric medium. 

Keeping in mind the result (\ref{linear}), the confinement hypothesis, 
which is illustrated in Fig. 2, can be understood as follows. Let us 
imagine that one tries to burst the meson up separating the quark from 
the antiquark. Stretching the meson requires pumping of the energy 
to the system. When the energy is sufficient to produce the quark-antiquark 
pair, the string breaks down and we have two mesons instead of one. 

At the end of this section we mention another commonly used model of the 
hadron structure the bag model. One assumes here that the vacuum exerts 
the pressure $B$ on the (colored) quarks and gluons. Then, the hadron is  
the bag with quarks and gluons similar to the bubble of vapour in the liquid. 
The bag is usually spherical but the deformations are possible. The model
appears to be really successful in describing a very reach mass spectrum
of hadrons.

\section{Quark-Gluon Plasma}

The quark-gluon plasma is the system of quarks and gluons, which are no 
longer confined in the hadron interiors, but can propagate in the whole 
volume occupied by the system. Thus, the quark-gluon plasma resembles 
the ionized atomic gas with quarks and gluons corresponding to electrons 
and ions and hadrons being the analogs of  atoms. One may wonder whether 
the existence of quark-gluon plasma is not in conflict with the confinement 
hypothesis. We note that the plasma is white as a whole. Thus, the color 
charges are still confined in the colorless system. However, it is still 
unclear why and when the hadrons can dissolve liberating quarks and gluons. 
We will consider this issue in the next section. Here we would like to 
discuss a somewhat unexpected consequence of the asymptotic freedom.

Due to the mentioned difficulties of the soft QCD, the properties of the 
hadronic matter - either composed of hadrons or of quarks and gluons - 
are poorly known at the moderate temperatures. The situation changes 
qualitatively when the system temperature $T$ is  much larger than the 
QCD scale parameter $\Lambda_{QCD}$. In this limit $T$ is the only 
dimensional parameter which describes the system. In particular, it 
determines the average momentum transfer in the interaction of quarks 
and gluons. Therefore, $\langle Q^2 \rangle = c\, T^2$ where $c$ is a 
dimensionless constant. On the basis of the dimensional argument we 
expect to achieve the asymptotic freedom regime (smallness of the 
coupling constant) when $T \gg \Lambda_{QCD}$. Then, the 
quark-gluon plasma is a weakly interacting gas of massless quarks 
and gluons. 

The detailed calculations performed within the thermal QCD show that 
the asymptotic freedom regime is indeed obtained in the high temeprature
limit. Here we will only add a simple physical argument to the dimensional 
one used above. The momentum transfer, which enters the formula of the 
running coupling constant (\ref{alpha}), corresponds (due to the Fourier 
transform) to the distance between the interacting partons. The distance 
is proportional to the inverse momentum transfer. When the plasma 
temperature increases, the density of the quark-gluon system grows as 
$T^3$ in the high temperature limit. (One can refer here again to the 
dimensional arguments or simple calculations presented in 
the next section.) Since the average particle separation in the gas of the 
density $\rho$ equals $\rho^{-1/3}$, the average inter-particle distance
decreases with $T$ as $T^{-1}$. Consequently, the average coupling 
constant vanishes when $T \rightarrow \infty $. 

\section{Deconfinement phase transition}

The transformation of the hadron gas into the quark-gluon plasma is called 
the deconfinement phase transition. There are many independent indications 
that such a transition indeed takes place in the dense hadronic medium. 
First of all one should mention the results obtained within the lattice 
formulation of QCD, where the space continuum is replaced by the discrete 
points. The Monte Carlo simulations show that there are two phases in the 
lattice QCD, which are identified with the hadron and quark-gluon phase, 
respectively. Here we would like to discuss simple arguments in favour 
of the existence of the quark-gluon plasma.

We start with the observation that hadrons are not point-like objects 
but their size is finite. The hadron radius is about 1 fm. So, let us 
consider the hadron gas which is so dense that the average separation 
of hadrons is about 1 fm. There is no reason to think  that the confining 
potential still operates in such a medium at the distances which are 
significantly larger than 1 fm.  When the quark and antiquark forming 
a meson are pulled way from each other there is no vacuum, which
expels the chromodynamic field, but there are other hadrons all around. 
Therefore,  the confining potential is expected to be screened at the
distances comparable to the average inter-particle separation.

There are two ways, as illustrated in Fig. 3,  to produce the dense 
hadronic matter. The first one is evident - squeezing of the nuclear matter. 
Since the baryon number, which is curried by neutrons and protons, is 
conserved, the nucleons cannot disappear and will start to overlap when 
the average inter-particle distance is smaller than the nucleon radius 
$r_N \cong 1$ fm. The inter-nucleon separation is smaller than $r_N$ 
at the densities exceeding 
$\rho = r_N^{-3} \cong 1 \; {\rm fm}^{-3} \cong 6 \: \rho_0$, 
where $\rho_0 = 0.16 \; {\rm fm}^{-3}$ is the so-called saturation 
or normal nuclear density being approximately equal the nucleon 
density in the nucleus center. 

The second method to produce the dense hadronic matter is to heat up 
the nuclear matter or the hadron gas.  The point is that in the contrast 
to the baryon or lepton number, which are the conserved quantities, 
the particle number is not. Therefore, when the gas temperature (measured 
in the energy units) becomes comparable to the particle mass, further 
heating leads not only to the increase of the average particle kinetic energy 
but to the growth of the average particle number. Of course, the particle 
number growth cannot violate the conservation laws. Therefore, the particles, 
which curry a conserved charge, must be produced in the particle-antiparticle 
pairs. For example, to keep the baryon number of the system fixed we can 
add to the system only pairs of the baryon and antibaryon. The essentially 
neutral particles such as $\gamma$ or $\pi^0$, which do not curry conserved 
charges, can be added to the system without restrictions, although the 
average particle number is controlled by the equilibrium conditions. More 
specifically, the numbers are determined by the minimum of the system free 
energy when the system temperature and volume are fixed.

Let us now consider the gas of noninteracting pions. Its temperature  
is assumed to be so high that the pions can be treated as massless particles. 
(The approximation appears to work quite well even for the temperatures 
which are close to the pion mass equal 140 MeV.) The pion density is then 
\begin{equation}\label{rho-pi}
\rho_{\pi} =  \int {d^3 p \over  (2\pi )^3} {g_{\pi} \over e^{E/T} -1 } 
= {g_{\pi} \zeta(3) \over \pi^2} \: T^3 \cong  0.73 \: T^3 \;,
\end{equation}
where $E \equiv |{\bf p}|$; $\zeta(z)$ is the Riemann function and 
$\zeta(3) \cong 1.202$; $g_{\pi}$ is the number of the particle internal 
degrees of freedom, which is 3 for the gas of  $\pi^+$, $\pi^0$ and 
$\pi^-$. One immediately finds from (\ref{rho-pi}) that the inter-pion 
separation is smaller than 1 fm for $T > 219$ MeV.

The deconfinement phase transition has been studied in the numerous 
phenomenological approaches. We consider here the simplest possible 
model, where the phase transition is assumed to be of the first order. 
Then, one can apply the Gibbs criterion to construct the phase diagram. 
The hadron phase is modeled by the ideal gas of  massless pions of three 
species ($g_{\pi}=3$) while the quark-gluon one by the ideal gas of 
quarks and gluons, which is baryonless i.e. the total baryon charge 
vanishes due to the equal number of quarks and antiquarks. Let us compute 
the number of the internal degrees of freedom in the quark-gluon gas. One 
should distinguish here the fermionic degrees of freedom of quarks $g_q$ 
and the bosonic of gluons $g_g$. There are two light quark flavours 
($u$ and $d$); there are quarks and antiquarks; we have two spin and 
three color quark states. So, $g_q = 2 \times 2 \times 2 \times 3 = 24$. 
The gluons are in two spin and eight color states, which give 
$g_g = 2 \times 8 = 16$.

Since the pressure of the ideal gas of massless particles equals one third
of the energy density, the pressure exerted by the pion gas is
\begin{equation}\label{press-pi}
p_{\pi} = {1 \over 3} \int {d^3 p \, E \over  (2\pi )^3}
{g_{\pi} \over e^{E/T} -1 } 
= {g_{\pi}  \pi^2 \over 90} \: T^4 \cong  0.33\: T^4 \;,
\end{equation}
while that of the quark-gluon plasma equals
\begin{equation}\label{press-qg}
p_{qg} = {1 \over 3} \int {d^3 p \, E\over  (2\pi )^3}
\bigg[ {g_g \over e^{E/T} -1 } + {g_q \over e^{E/T} + 1 }\bigg]
= \Big(g_g + {7 \over 8} g_q \Big)\, {\pi^2 \over 90} \: T^4 
\cong  4.1\: T^4 \;.
\end{equation}

According to the Gibbs criterion the phase, which generates the higher 
pressure at a given temperature, is realized. Then, one finds from
eqs. (\ref{press-pi}, \ref{press-qg}) that the pressure of the quark-gluon
plasma is always greater than that of pions. Therefore, we should have,
in conflict with the experiment, the quark-gluon phase at any temperature. 
However, we have not taken into account the pressure exerted by the 
vacuum on quarks and gluons. Subtracting the bag constant $B$ from the 
r.h.s. of eq. (\ref{press-qg}), one finds that below the critical temperature 
$T_c$ there is the pion gas and above the quark-gluon plasma. The critical 
temperature is given as
$$
T_c = \Bigg[{ 90 B \over \pi^2 \big( g_g + {7 \over 8}g_q 
- g_{\pi}\big)} \Bigg]^{1/4} \cong 0.72 \:B^{1/4} \;.
$$
Taking $B^{1/4} = 200$ MeV, we get $T_c = 144$ MeV.

In Fig. 4 we show a schematic phase diagram of the strongly interacting 
matter. The baryon density is measured in the units of the normal nuclear
one. The point at $\rho = \rho_0$ and $T=0$ represents normal nuclei.
In fact, this is the only point in the diagram which is really well known
and understood. At $ \rho > \rho_0$ and $T = 0$ there is a region of
the dense nuclear matter with several exotic forms being suggested. 
When the baryon density exceeds $2 - 3 \rho_0$ one expects 
a transition to the quark-gluon phase. At even higher densities the plasma
is believed to be perturbative i.e. weakly interacting. So, one deals 
with the quasi ideal strongly degenerated quark gas. The nuclear matter at 
the temperature larger than a few MeV is traditionally called a hadron gas 
being mostly composed of pions when the baryon density vanishes. Then, 
we have quite reliable QCD lattice results which tell us that there is the 
deconfinement phase transition at $T \cong 180$ MeV. With the significantly 
larger temperatures we approach again the perturbative regime where the 
plasma is weakly interacting. In the next section we discuss how the phase 
diagram can be explored by means of the heavy-ion collisions.

\section{Heavy-Ion Collisions}

As already mentioned in the introduction, the heavy-ion collisions provide 
a unique opportunity to study the quark-gluon plasma in the laboratory 
experiments. More precisely, a drop of a dense hadronic matter can be
created in the collisions; the question whether the matter is for some
time in the deconfined phase is a separate issue. 

The physics of heavy-ion collisions essentially depends on the collisions 
energy. A good measure is not the whole energy of the incoming nucleus 
but the energy per nucleon. The point is that at the energy of a few GeV 
per nucleon, which is the lowest energy being interesting from the point 
of view of the quark-gluon plasma, the nucleus does not interact 
as a whole but there is an interaction of the overlapping parts of the 
colliding nuclei as depicted in Fig. 5. The nucleons to be found in these 
parts are called the {\it participants} while the remaining ones the 
{\it spectators}.  One often distinguishes the central from the peripheral 
collisions. In the latter ones, which proceed with a large value of the 
impact parameter, most of the nucleons are spectators. In the case of the 
central collisions basically all nucleons from the smaller nucleus (target
or projectile) are participants and the interaction zone is the largest. 
Obviously the central collisions are the most interesting in the context 
of the quark-gluon plasma searches. Unfortunately, the nucleus-nucleus 
cross section is dominated by the peripheral collisions. The cross section 
contribution of the collisions with a given impact parameter 
$b$ is $2 \pi b db$. Therefore, the contribution vanishes when 
$b \rightarrow 0$.

The high-energy nucleus-nucleus collision proceeds according to the 
following scenario. The overlapping parts of the colliding nuclei 
strongly interact and a dense and hot hadronic system, which is often 
called a {\it fireball}, is created. Initially it is formed either 
by the quarks and gluons or by the hadrons. Then, the system expands 
and cools down. If the fireball matter has been initially in the deconfined 
phase, the system experiences the hadronization i.e. the quarks and 
gluons are converted into the hadrons. The fireball further expands.
When its density is so low that the mean free path of the hadrons 
is close to the system size or the expansion velocity is comparable
the that of the individual particles, the whole system decouples into 
the hadrons which do not interact with each other any longer. 
However, the unstable particles can still gradually decay into the final
state hadrons. The moment of decoupling is called the {\it freeze-out}, 
because the particle momenta  are basically fixed at that time. Thus, 
the final state hadrons, which are observed by experimentalists in 
the particle detectors, characterize the fireball at the freeze-out.

As we discussed in Sec. V, the dense hadronic matter can be obtained 
either due to the nuclear matter compression or the nuclear matter heating. 
It appears that the significant effect of compression at the early stage of
heavy-ion collisions, which is still insufficient for the quark-gluon 
generation, is observed at the relatively low energies, not larger than 
1 GeV. Then, one can study the dense nuclear matter of rather small 
temperature. At higher energies the atomic nuclei appear to be rather 
transparent i. e. the colliding nuclei traverse each other. The participants 
are only slightly deflected from their straight line trajectories due to 
the interaction. However, they lose a sizable portion of their energy which 
is further manifested in the form of produced particles, mainly pions. 
Therefore, the baryon density of the produced hadronic system is not much 
increased even at the early collisions stage. When the produced system 
expands, the baryon density soon gets a value which is smaller than the 
normal nuclear density. The strongly interacting matter however is 
significantly heated up. If the temperature exceeds the deconfinement 
transition temperature, the matter is expected to be in the quark-gluon 
phase. 

A comment is in order here. We use the concept of the temperature which 
explicitly assumes that the system is in the thermodynamic equilibrium. 
It is far not obvious that this is really the case. Any physical system needs 
some time to reach the state of equilibrium. The hadronic matter produced 
in heavy-ion collisions cannot achieve the {\it global} equilibrium but the 
theoretical as well as experimental arguments suggest that the {\it local} 
quasi equilibrium is possible. The global equilibrium is characterized by 
the thermodynamic parameters which are unique for the whole system. 
Before the global equilibrium is achieved a system is usually for some time 
in the local equilibrium state. The system's parts are then already in the 
equilibrium but the thermodynamic parameters - temperature, density, 
hydrodynamic velocity - vary from part to part. The hadronic matter produced 
in heavy-ion collisions is however not kept in any container but it immediately 
starts to expand, mainly along the beam axis. Consequently there is a sizable 
variation of the hydrodynamic velocity in this direction.

\section{Plasma Signatures}

The plasma creation is expected to occur at the early collision stage, but 
it must hadronize i.e. experience the transition to the hadron gas, when 
the matter is expanding and cooling down at the later stages. Therefore, 
we always observe hadrons in the final state of the collisions and it is 
really difficult to judge whether the plasma has been present or not. 
Although it is hard to imagine a smoking gun proof, a few signatures of 
the plasma creation has been proposed. We discuss below the two which
seem to be the most promising.

It has been argued that the presence of the plasma at the early collision
stage increases the number of strange particles observed in the final state.
The point is that the mass of strange quark mass appears to be significantly
smaller than that of strange particles. The strange $(s)$ quark mass is 
about 150 MeV. Therefore, one needs 300 MeV to produce the $s-\bar s$
pair. The strange quarks must be, of course, produced in pairs because the 
strangeness is conserved in the strong interactions. The most energetically
favorable way to produce the strangeness at the hadron level proceeds
in the reaction $\pi + N \rightarrow K + \Lambda $. Here the incoming
particles - the pion and nucleon - must curry over 500 MeV in the center 
of mass frame. Thus, it is easier to produce strangeness at the quark-gluon 
than hadron level. Once the strange quarks appear in the plasma, they cannot 
disappear - a rather rare annihilation process of $s-\bar s$ pairs can be 
neglected - and consequently, they are distributed among the final state hadrons. 
Quantitative comparison of the two scenarios with and without the 
plasma indeed shows that the presence of the plasma leads to the 
significant strangeness enhancement.

The second plasma signatures deals with the $J/\psi$ particle or 
charmonium which is a bound state of  the charm quark $c$ and  
antiquark $\bar c$. The $J/\psi$ particle is expected to dissolve
much easier in the quark-gluon environment than in the hadron one. 
This can be understood as a result screening of the potential, which 
binds the $c$ and $\bar c$ quarks, by the color charges of the plasma 
particles. Therefore, the number of the $J/\psi$ particles in the final 
state should be significantly reduced if the plasma is present at the 
collision early stage.

The two predicted quark-gluon plasma signatures have been indeed 
experimentally observed in the central heavy-ion collisions at the 
energy of about 200 GeV per nucleon which have been recently 
studied at CERN. The whole set of the experimental data however 
does not fit to the theoretical expectations. There have been 
advocated the mechanisms of the strangeness enhancement 
and $J/\psi$ suppression which are different than those mentioned 
above. Therefore, it is a matter of hot debate whether the plasma 
is produced at the energies which currently available. The plasma 
generation at higher energies seems to be guaranteed.

\section{Perspectives}

In the near future the nucleus-nucleus collisions will be studied at the 
accelerators of a new generation: Relativistic Heavy-Ion Collider (RHIC) 
at Brookhaven and Large Hadron Collider (LHC) at CERN. The collision 
energy will be larger by one or two orders of magnitude than that of 
the currently operating machines. The heavy-ion experiments have been
performed till now in the beam-targets system where the accelerated ion 
is smashed against the target nucleus which is initially at rest. RHIC
and LHC will use another principle - there will be accelerated two 
intersecting ion beams. The energy of each beam will be 100 GeV 
per nucleon at RHIC and 3 000 GeV at LHC. Thus, the collision 
energy in the center of mass frame will equal, respectively, 200 and 
6 000 GeV which should be compared to 20 GeV presently available 
in the beam-target systems.

The proton-proton collisions have been already studied experimentally
at the energy domain of RHIC. The perturbative QCD, which is hardly 
available at lower energies, is extensively used here. This is possible 
because the average momentum transfer grows with the collision 
energy and the QCD coupling constant is ten relatively small. Therefore, 
the theoretical understanding of the collisions, paradoxically, improves 
with the growing energy.

The nuclear collisions at RHIC or LHC are expected to be so violent 
that the quarks and gluons comprising the nucleons will be easily 
deconfined, for some time of course. At such huge energies the nucleon 
is visualized as a cloud of partons which breaks up into the parton 
showers as a result of collision with another nucleon. Thus, the creation 
of the quark-gluon plasma seems to be unavoidable at RHIC and LHC.
The method of its detection however remains to a large extend an open 
question.

\vspace{1cm}

I am very grateful to Marek Ga\'zdzicki for critical reading of the 
manuscript.

\vspace{1cm}
\begin{center}
{\bf Figure Captions}
\end{center}
\vspace{0.5cm}
\noindent

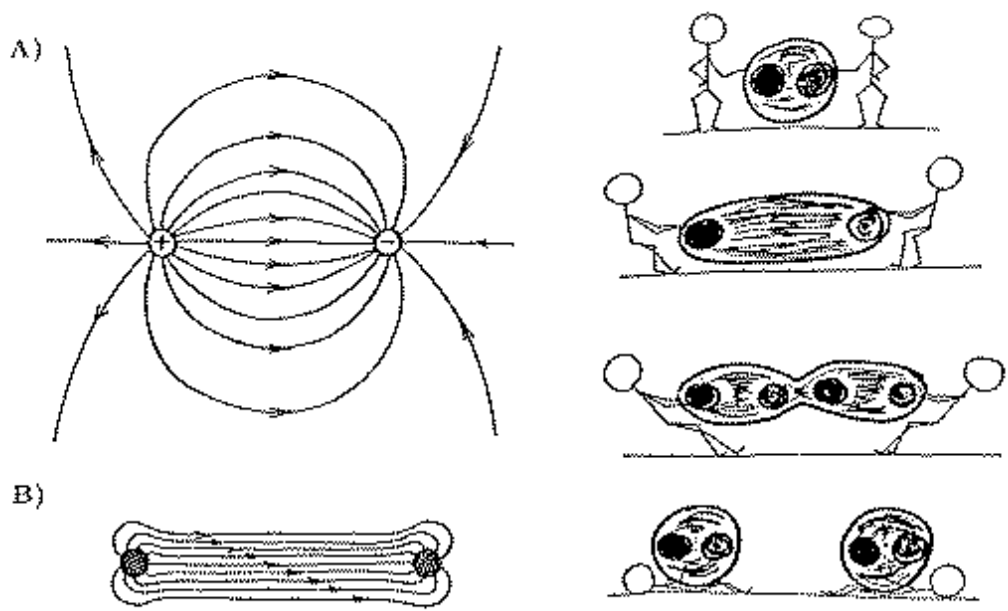
{\bf Fig. 1.} 
The electric field lines in the vacuum (a) and diaelectric medium (b).

\vspace{0.5cm}
\noindent

{\bf Fig. 2.} 
The confinement.

\vspace{0.5cm}
\noindent

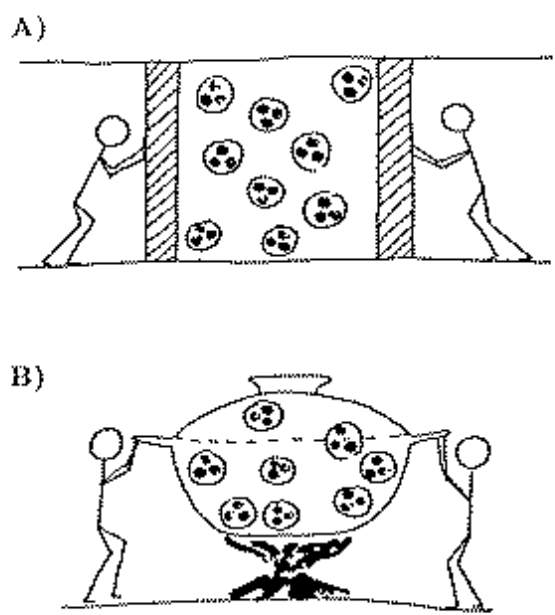
{\bf Fig. 3.} 
Two technologies to produce the dense hadronic matter: compression (a)
and heating (b).

\vspace{0.5cm}
\noindent

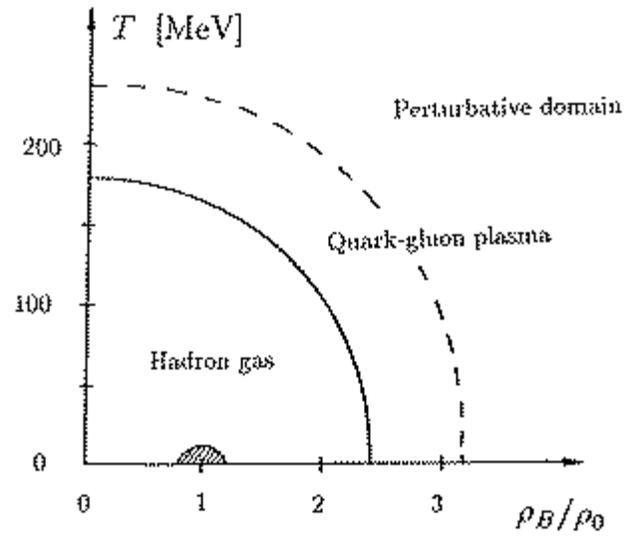
{\bf Fig. 4.} 
The phase diagram of the strongly interacting matter.

\vspace{0.5cm}
\noindent

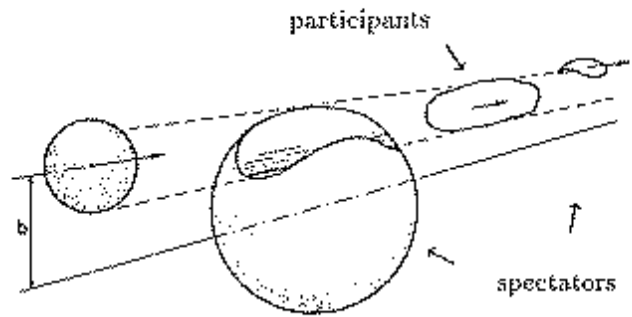
{\bf Fig. 5.} 
The geometry of the nucleus-nucleus collision at high energy.

\end{document}